\documentclass[lettersize,journal]{IEEEtran}
\usepackage{amsmath,amsfonts}
\usepackage{algorithmic}
\usepackage{algorithm}
\usepackage{array}
\usepackage[caption=false,font=normalsize,labelfont=sf,textfont=sf]{subfig}
\usepackage{textcomp}
\usepackage{stfloats}
\usepackage{url}
\usepackage{verbatim}
\usepackage{graphicx}
\usepackage{cite}
\usepackage{amsmath, amsfonts, amssymb, amsthm, bm,graphicx,multirow, cite, stfloats}
\usepackage{array, enumerate, diagbox, subcaption}
\usepackage{color}
\usepackage{algorithm, algorithmic,setspace}

\usepackage{soul}

\begin{document}

\title{A Block Quantum Genetic Interference Mitigation Algorithm for Dynamic Metasurface Antennas \\ and Field Trials}

\author{Taorui Yang,
Haifan Yin,~\IEEEmembership{Senior Member,~IEEE}, Rongguang Song, Lianjie Zhang 
\thanks{T. Yang, H. Yin, R. Song, and L. Zhang are with the School of Electronic Information and Communications, Huazhong University of Science and Technology,
Wuhan 430074, China. E-mail: \{try, yin, song\_rg, lianjie\}@hust.edu.cn. The corresponding author is Haifan Yin.}
\thanks{This work was supported by the Fundamental Research Funds for the Central Universities and the National Natural Science Foundation of China under Grant 62071191.}
}

\maketitle

\begin{abstract}

This paper proposes a quantum algorithm for Dynamic Metasurface Antennas (DMA) beamforming to suppress interference for an amplify-and-forward relay system in multi-base station environments.
This algorithm introduces an efficient dynamic block initialization and overarching block update strategy, which can enhance the Signal-to-Interference-plus-Noise Ratio (SINR) of the target base station (BS) signal without any channel information.
Furthermore, we built a relay system with DMA as the receiving antenna and conducted outdoor 5G BS interference suppression tests. 
To the best of our knowledge, this is the first paper to experiment DMA in commercial 5G networks. 
The field trial results indicate an SINR improvement of over 10 dB for the signal of the desired BS. 

\end{abstract}


\section{Introduction}

\IEEEPARstart{D}{ynamic} metasurface antennas (DMA), as an emerging future antenna array technology based on reconfigurable metamaterials, demonstrates promising prospects in the next-generation wireless communications \cite{DMA_for_6G}.
Unlike reconfigurable intelligent surfaces (RIS), another metamaterial-based technique that modifies reflection/refraction characteristics to reshape the wireless propagation environment, DMA allows for programmable control of the transmit and receive beam patterns directly \cite{Metantenna}.
Compared with the traditional hybrid architectures, DMA can achieve similar performance with lower hardware complexity, decreased power consumption, and reduced costs\cite{Energy-EfficientMISORobertW,DMA_Uplink_Massive_MIMO2019}.
Researchers have explored the application of DMA in multi-user communication.
The work in \cite{DMA_Uplink_Massive_MIMO2019} designed an alternating configuration algorithm for the uplink multi-user communication, while the authors in \cite{Downlink_Beamforming2023} improved the Signal-to-Noise Ratio (SNR) and weighted sum rate in the downlink scenario. The authors in \cite{Beamforming_Design_MU_MISO_Downlink2024} devised a Cross-Entropy Optimization-based algorithm to address the issue of discrete DMA coefficients.

However, the studies mentioned above are all based on the assumption that channel information is known.
In engineering practice, the channel estimation problem is particularly complex due to the large number of antenna elements but significantly fewer RF chains in the DMA, which leads to high pilot overhead.
Additionally, when applied in amplify-and-forward (AF) relay systems, the signals are not decoded due to a lack of RF chain, making the acquisition of channel information even more challenging.
To address this practical issue, we propose a dynamic block quantum genetic (DBQG) algorithm that only utilizes limited received signal indicators from the user side, which does not require any prior channel knowledge or collaboration from the BS side.

To optimize the discrete phase shift coefficients of DMA using only the signal quality metrics of the received signals, evolutionary algorithms such as genetic algorithm (GA) are available  \cite{GA_RIS}. 
Quantum genetic algorithms (QGA) incorporate principles of quantum computing into GA, significantly accelerating the convergence rates\cite{QGA_2008}. 
However, it is susceptible to local optima and suffers from premature convergence issues. 
To address this issue, we designed a diverse adaptive quantum updating strategy and incorporated the characteristics of the optimization problems for the DMA array in DBQG to enhance performance and convergence speed.

Our simulation results demonstrate the capability of the proposed DBQG algorithm to configure DMA for spatial interference filtering, as well as its excellent convergence speed and ability to overcome local optima.
To the best of our knowledge, few tests have been conducted on DMA in practical wireless networks. 
We validated the algorithm in a self-built DMA relay system and conducted interference suppression tests in commercial 5G networks of China Mobile.
Our experimental results demonstrate the excellent performance of DBQG in addressing this practical communication problem, thereby validating the capability of our method in mitigating inter-cell interference.

\section{SYSTEM MODEL}
\label{sec 2}

\begin{figure}[htbp]
\centerline{\includegraphics[width = .64\linewidth]{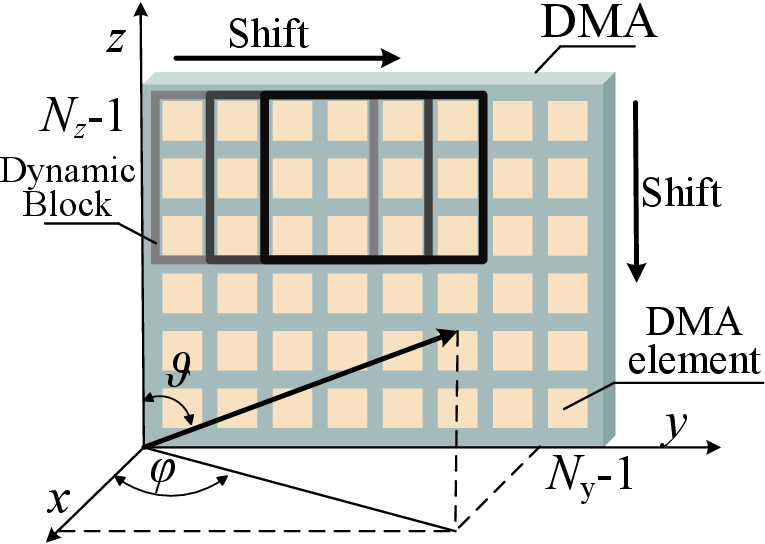}}
\caption{System model and dynamic block shifting.}
\label{fig:sliding_window}
\end{figure}

We consider a typical downlink communication system with $K$ BSs equipped with $M  = M_y M_z$ antennas and a relay with
DMA shown in Fig.\ref{fig:sliding_window}. While the serial-fed architectures of most DMAs generally result in phase-amplitude coupling, our DMA with $N = N_y N_z$ parallel-fed elements exhibits weak coupling, which is regarded as a non-ideal factor.
In engineering applications, the phase shift of the DMA elements is usually quantized into discrete bits.
The phase shift of a $\tau$-bit quantized DMA may take $2^\tau$ values, denoted by $\Phi_{\tau} = \{0, \frac{2\pi}{2^\tau}, \frac{2 \cdot 2\pi}{2^\tau}, \ldots, \frac{(2^\tau - 1) \cdot 2\pi}{2^\tau}\}$. 
The phase shift coefficients of the DMA are denoted by $\bm{\theta} = \begin{bmatrix} e^{j\theta_{1}}, e^{j\theta_{2}}, \ldots, e^{j\theta_{N}}\end{bmatrix} \in \mathbb{C}^{1 \times N}$, with $\theta_{n} \in \Phi_{\tau}$ representing the phase shift of the $n$-th element. 

Let ${\bf H}_{k} \in \mathbb{C}^{N \times M}$ denote the channel between the $k$-th BS 
\clearpage
\noindent{and the DMA:}
\begin{equation}
{\bf H}_{k}=\sum_{l=1}^{L_k}\beta_{k}^{l}\boldsymbol{\alpha}_{r}^{\mathrm{T}}(\vartheta_{k,r}^{l},\varphi_{k,r}^{l})  \boldsymbol{\alpha}_{BS}(\vartheta_{k,BS}^{l},\varphi_{k,BS}^{l}) ,
\end{equation}
where $L_k$ is the number of paths between the DMA and the $k$-th BS, $\beta_{k}^{l}$ is the complex path gain of the $l$-th path, $\vartheta_{k,r}^{l}$ ($\varphi_{k,r}^{l}$) and $\vartheta_{k,BS}^{l}$ ($\varphi_{k,BS}^{l}$) denote the elevation (azimuth) angle at the DMA and the antenna of BS. 
Moreover, $\boldsymbol{\alpha}_{r}(\vartheta,\varphi)$ and $\boldsymbol{\alpha}_{BS}(\vartheta,\varphi)$ represents the steering vector for the DMA and the antenna of BS. For the DMA, it can be represented as
\small
\begin{equation}
\begin{aligned}
{\boldsymbol\alpha}_{r}(\vartheta, \varphi) &= \begin{bmatrix} 
(\alpha_y(\vartheta,\varphi))^{0}, (\alpha_y(\vartheta,\varphi))^{1}, \ldots, (\alpha_y(\vartheta,\varphi))^{N_y-1}
\end{bmatrix}\\
& \otimes \begin{bmatrix} 
(\alpha_z(\vartheta,\varphi))^{0}, (\alpha_z(\vartheta,\varphi))^{1}, \ldots, (\alpha_z(\vartheta,\varphi))^{N_z-1}
\end{bmatrix}.
\end{aligned}
\end{equation}
\normalsize
$\alpha_y(\vartheta,\varphi)$ and $\alpha_z(\vartheta,\varphi)$ denote the spatial signatures:
\begin{equation}
\begin{aligned}
\alpha_y(\vartheta,\varphi) &= \text{exp}(-j2\pi\frac{d_y}{\lambda}\sin(\vartheta)\sin(\varphi)), \\
\end{aligned}
\end{equation}
\begin{equation}
\begin{aligned}
\alpha_z(\vartheta,\varphi) &= \text{exp}(-j2\pi\frac{d_z}{\lambda}\cos(\varphi)),
\end{aligned}
\end{equation}
where $d_y$ and $d_z$ are the element spacings along the $y$-axis and $z$-axis, respectively, and $\lambda$ is the freespace wavelength.

Hybrid beamforming techniques are commonly employed in current 5G macro BSs with 8 fixed analog beams corresponding to different angular directions. 
We denote the corresponding analog beamforming vector autonomously selected by the BS as ${\bf w}_{k} \in \mathbb{C}^{M \times 1}$.
Let $s_{k}$ denote the downlink transmitted signal from the $k$-th BS. 
We consider the signal $s_{t}$ from the $t$-th BS as the desired signal, while the signals from other BSs are interference. The received signal is expressed as 

\begin{equation}
 y =  \bm{\theta}_{t} {\bf H}_{t} {\bf w}_{t} s_{t} + \sum_{k \neq t}^{K} \bm{\theta}_{t} {\bf H}_{k} {\bf w}_{k}  s_{k} + n,
\end{equation}
where $n \sim \mathcal{CN}\left(0, \sigma^{2}\right)$ represents the additive Gaussian noise (AWGN) at the receiver and $\bm{\theta}_{t}$ is the phase shift coefficients on the DMA for the desired signal. 

Our objective is to suppress interference at the receiver and maximize the received Signal-to-Interference-plus-Noise Ratio (SINR).  
The problem is formulated as
\begin{equation}
\begin{aligned}
&\max_{\bm{\theta}_{t}} &\frac{\left| \bm{\theta}_{t} {\bf H}_{t} {\bf w}_{t} \right|^{2} p_{t}}
{\sum_{k \neq t}^{K}\left| \bm{\theta}_{t} {\bf H}_{k} {\bf w}_{k} \right|^{2} p_{k}+\sigma^{2}}\\
&\mathrm{s.t.} & \theta_{n} \in \Phi_{\tau}, \forall n = 1, \ldots, N,
\end{aligned}
\end{equation}
where $p_{k}$ is the transmit power of the $k$-th BS.

The challenge of solving the above problem is two-fold. 
Firstly, from a practical perspective, acquiring channel information in an AF relay system is highly challenging. 
Secondly, both the quadratic objective function and the discrete phase shift constraints are non-convex.
To address this issue, we propose a blind beamforming algorithm inspired by QGA, which will be introduced in the subsequent sections.

\section{Dynamic Block Quantum Genetic Algorithm}
\label{sec 3}

Quantum Genetic Algorithm (QGA) combines principles of quantum computing and genetic algorithm to enhance optimization processes. In QGA, quantum bits are used to represent individuals, which can superimpose multiple possibilities and enable parallelism and faster convergence instead of using deterministic binary bits. The quantum bit state of a gene is expressed as $\left| \psi \right\rangle = \gamma \left| 0 \right\rangle + \delta \left| 1 \right\rangle$, where $\gamma$ and $\delta$ denote the complex probability amplitudes of the corresponding states and satisfy ${\left| \gamma \right|^2} + {\left| \delta \right|^2} = 1$. And, $\left| 0 \right\rangle $ and $ \left| 1 \right\rangle$ represent quantum states ``0" and ``1" respectively. Through the collapse operation, the quantum bits will be measured based on probability amplitude and reach definite quantum states ``0" or ``1" to be used as potential solutions to the problem.

After the collapse of individuals, the obtained solutions are then applied to the target problem to determine the fitness of each individual. The population is typically updated by altering the probability amplitudes of individuals using quantum rotation gates, and the process can be represented as
\begin{equation}
\begin{bmatrix}\alpha_{i}^{\prime} \\ \beta_{i}^{\prime}
\end{bmatrix} =
\begin{bmatrix}
\cos(\triangle\omega_{i}) & -\sin(\triangle\omega_{i}) \\
\sin(\triangle\omega_{i}) & \cos(\triangle\omega_{i})
\end{bmatrix}
\begin{bmatrix} \alpha_{i} \\ \beta_{i} \end{bmatrix},
\end{equation} 
where $(\alpha_{i},\beta_{i})$ and $(\alpha_{i}^{\prime},\beta_{i}^{\prime})$ represent the probability amplitudes of the $i$-th quantum bit of the pre-update and post-update individual, respectively. The rotation angle $\triangle\omega_{i}$ is related to the fitness of the individuals and plays a crucial role in determining the performance and convergence speed.

By exploiting the cyclic offset property between neighboring DMA elements and the proximity quantization property of the DMA phase shift coefficients, we propose an efficient dynamic block initialization and updating strategy in our DBQG method.

\subsection{Initialization Incorporating DMA Characteristics}
\label{sec_3_A}

The probability amplitudes in QGA are usually initialized to the same value, resulting in a slow convergence speed. 
In DBQG, based on the cyclic offset property and the proximity quantization property between neighboring elements of DMA optimization, a dynamic block method is designed to efficiently obtain higher initial SINR values. 

Focusing on the LoS path of the desired signal, the gain of the effective channel is expressed as
\begin{equation}
\begin{aligned}
\left| \bm{\theta}_{t} {\bf H}_{t} {\bf w}_{t} \right|^{2} 
&= \left| \beta_{t} \bm{\theta}_{t}   
\boldsymbol{\alpha}_{t,r}^{\mathrm{T}}(\vartheta_{t,r},\varphi_{t,r}) \boldsymbol{\alpha}_{t,BS}(\vartheta_{t,BS},\varphi_{t,BS}) {\bf w}_{t}  \right|^{2} \\
&= \left| \eta_{t} \right|^{2} \left| a_{1} e^{j\theta_{1}} + a_{2} e^{j\theta_{2}} + \ldots + a_{N} e^{j\theta_{N}} \right|^{2},
\end{aligned}
\end{equation}
where 
\begin{align} 
a_{n} &= e^{\frac{-j2\pi}{\lambda} (d_y \sin(\vartheta_{t,r})\sin(\varphi_{t,r}) y_n + d_z \cos(\varphi_{t,r}) z_n)}, \\
\eta_{t} &= \beta_{t} \boldsymbol{\alpha}_{t,BS}(\vartheta_{t,BS},\varphi_{t,BS}) {\bf w}_{t} .
\end{align} 
Further, expand it as
\small
\begin{equation}
\begin{aligned}
\left| \bm{\theta}_{t} {\bf H}_{t} {\bf w}_{t} \right|^{2} 
&= N \left| \eta_{t}\right|^{2} + 2 \sum_{i=1}^{N-1}\sum_{j=i+1}^{N} \left| \eta_{t}\right|^{2} \cos(\rho_{ij}(\vartheta_{t,r},\varphi_{t,r})),
\end{aligned}
\end{equation}
\normalsize
with
\begin{equation}
\begin{aligned}
\rho_{ij}(\vartheta_{t,r},\varphi_{t,r}) &= \frac{-2\pi}{\lambda}(d_y (y_i - y_j) \sin (\vartheta_{t,r})\sin (\varphi_{t,r}) \\
&+ d_z (z_i - z_j) \cos (\varphi_{t,r})) + (\theta_{i} - \theta_{j}),
\end{aligned}
\end{equation}
where $ y_i \in \{0,\ldots,N_y-1 \} $ and $ z_i \in \{ 0,\ldots,N_z-1 \} $ represent the coordinates of the $i$-th element in the y-axis and z-axis directions, respectively. 

Therefore, the condition for maximizing the channel gain is that for any two elements $i$ and $j$, their phase shifts 
satisfy:
\begin{equation}
\begin{aligned}
\theta_{i} - \theta_{j} &= 2 f \pi + \frac{2\pi}{\lambda}(d_y (y_i - y_j) \sin (\vartheta_{t,r}) \sin (\varphi_{t,r}) \\
&+ d_z (z_i - z_j) \cos (\varphi_{t,r}) ), f \in \mathbb{Z}. 
\end{aligned}
\end{equation}

In the y-axis and z-axis direction, the phase shift differences between adjacent elements are $ 2f\pi + \frac{2\pi}{\lambda} d_y \sin (\vartheta_{t,r})\sin (\varphi_{t,r}) $ and $ 2f\pi + \frac{2\pi}{\lambda} d_z \cos (\varphi_{t,r}) $, respectively. 
Utilizing this cyclic offset property, we iteratively apply the cyclic phase offsets to the DMA elements one by one to enhance channel gain.
Additionally, the optimal phase coefficients of neighboring elements usually differ in the continuous domain, yet they are sometimes quantized to identical values. 
Therefore, by utilizing this proximity quantization property, we simultaneously perform the offset operation on an entire block of elements to further enhance algorithm efficiency.

Based on the analysis above, we propose a dynamic block initialization strategy, and the main procedure is as follows:

\begin{itemize}
    \item The elements within a range of $\frac{N_y}{b}$ columns and $\frac{N_z}{b}$ rows are considered as a dynamic block where $b$ is the block size parameter. The top-left vertex of the initial block is located at $(0, N_{z}-1)$.
    
    \item Sequentially apply cyclic phase offset $\triangle {\theta}_{i} \in \Phi_{\tau}$ on top of the original phase shift coefficients to all elements within the block, where $i = 0, \ldots, 2^{\tau}-1$. Record the corresponding signal indicator $I_i$. 
    The offsets are applied to elements within the block integrally, and the differences in their phase shift coefficients will be preserved.
    
    \item Select the phase offset $\triangle {\theta}_{\text{opti}}$ that yields the highest signal indicator $I_{\text{opti}}$ and apply it to elements within the block.
    
    \item Shift the block range 
    as a whole 
    one column to the right and repeat the operation of finding and applying the optimal 
    phase 
    offset.
    
    \item After updating one row, shift the block range downwards and continue the operation from the leftmost position until the entire DMA array is updated.
\end{itemize}

The shifting procedure is shown in Fig.\ref{fig:sliding_window}. 
This initialization strategy may lead to a relatively better solution with
computational complexity ($\mathcal{O}(2^{\tau}\frac{(b-1)^{2}}{b^2} N_yN_z)$) and significantly enhance the convergence speed of the DBQG.

\subsection{Diverse Quantum Update Strategy}

The rotation angle in QGA is determined based on a lookup table with fixed discrete values, making the search process relatively inefficient and time-consuming. A diverse adaptive update strategy was introduced in DBQG to overcome this limitation.
This strategy adjusts the evolution speed of individuals based on their similarities and relative signal indicators.

We utilize the Hamming distance to evaluate the similarity between two deterministic solutions. 
After applying all individual solutions to the DMA and obtaining the corresponding signal indicators, the Hamming distances between the individuals are computed. 
Let $D_{q}^{g}$ represent the Hamming distance between the $q$-th individual and the individual with currently the highest SINR in the $g$-th generation. The similarity between the $q$-th individual and the current best individual is then represented as
\begin{equation}
S_{q}^{g} =
\begin{cases}
\frac{D_{q}^{g} - D_{\text{min}}^{g}}{D_{\text{max}}^{g} - D_{\text{min}}^{g}} &  D_{\text{max}}^{g} \neq D_{\text{min}}^{g} \\
0 & D_{\text{max}}^{g} = D_{\text{min}}^{g} \\
\end{cases},
\label{similarity_idv}
\end{equation}
where $D_{\text{max}}^{g}$ and $D_{\text{min}}^{g}$ represent the maximum and minimum Hamming distances, respectively, between any individual and the current best individual. 

Let $S_{pop}^{g}$ denote the average of the similarity of all individuals in the $g$-th generation. 
The similarity of individuals and the entire population will affect subsequent operations.

In our algorithm, the rotation angle of the $q$-th individual is calculated as
\begin{equation}
\triangle \omega_{q}^{g} = \triangle\omega_{0} \cdot \exp(a \frac{I_{\text{best}}^{g} - I_{q}^{g}}{I_{\text{best}}^{g}} + (a - 1) \cdot S_{q}^{g}),
\label{rotation_angle}
\end{equation}
where $\triangle\omega_{0}$ is the rotation angle step size, and the weighting factor $a \in (0,1)$ determines the influence of both the relative signal indicator values of individuals and their similarity to the optimal individual on individual evolution.
Additionally, $I_{q}^{g}$ and $I_{max}^{g}$ represent the signal indicator value of the $q$-th individual in the $g$-th generation and the current maximum signal indicator value, respectively.

To address the issue of local convergence in QGA, DBQG introduces a block mutation operation. 
Mutation operations typically involve randomly altering a bit of an individual to maintain population diversity.
For DMA, simply changing the phase shift coefficients of one or two elements often does not significantly alter the overall beamforming performance. Therefore, we will perform extensive mutation operations within randomly selected blocks. The procedure is as follows:

\begin{itemize}
    \item Select individuals for mutation based on the mutation probability $P_{\text{muta}}$, which is primarily derived from practical engineering experience and determined by the current similarity of the population and individuals, as follows:
    \begin{equation}
    P_{\text{muta}} = P_{0} \cdot (1 - S_{pop}^{g}) \cdot \exp{(1 - S_{q}^{g})},
    \label{mutation_probability}
    \end{equation}
    where $P_{0}$ is the predefined basic mutation probability.
    \item For the selected individuals, the elements within a mutation block are chosen as the target elements. The scope and position of the mutation block are randomly selected.
    \item For each quantum bit corresponding to each element in the mutation block, a Hadamard gate mutation is applied. 
\end{itemize}

The detailed procedure of DBQG is shown in Algorithm \ref{alg:1}.
The DBQG preserves the optimal individual during the iteration process, ensuring a monotonic increase in the objective function value. It can be proven to converge with probability $1$ utilizing the ergodicity and stationary distribution of a finite-state homogeneous Markov chain.
Due to the absence of complex matrix operations, our algorithm has relatively low computational complexity ($\mathcal{O}(GT)$, where $G$ is the population size, and $T$ is the number of generations). 

\begin{algorithm}[ht]
  \renewcommand{\algorithmicrequire}{\textbf{Input:}}
  \renewcommand{\algorithmicensure}{\textbf{Output:}}
  \caption{Dynamic Block Quantum Genetic Algorithm}
  \label{alg:1}
  \begin{algorithmic}[1]
    \REQUIRE The maximum  number of iterations $T_{\text{max}}$, rotation angle step size $\triangle\omega_{0}$, basic mutation probability $P_{0}$.
    \ENSURE Phase shift coefficients $\bm{\theta}_{t}$.

    \STATE Execute the dynamic block initialization according to \ref{sec_3_A}:
    \REPEAT
        \STATE Apply phase offsets $\triangle {\theta}_{i}$ to the block iteratively;
        \STATE Find the optimal offset $\triangle {\theta}_{\text{opti}}$ and apply it to the block;
        \STATE Move the dynamic block;
    \UNTIL {the entire DMA array is updated.}
    \STATE Add the result to the initial population and randomly initialize other individuals.
    \STATE Set $g \gets 0$.
    \WHILE{$g < T_{max}$}
        \STATE Collapse the population and obtain $\bm{\theta}_{t,q}^{g}$;
        \STATE Apply all $\bm{\theta}_{t,q}^{g}$ to DMA and obtain the signal indicator $I_{q}^{g}$;
        \STATE Update the optimal individual $\bm{\theta}_{t,\text{best}}^{g}$;
        \STATE Calculate the similarities $S_{q}^{g}$ and $S_{p}^{g}$ according to (\ref{similarity_idv});
        \STATE Calculate $\triangle \omega_{q}^{g}$ according to (\ref{rotation_angle}), and perform rotation gate operations;
        \STATE Execute block mutation operation according to (\ref{mutation_probability});
        \STATE Set $g \gets g + 1$;
    \ENDWHILE
    \STATE \textbf{return} The optimal coefficients $\bm{\theta}_{t, \text{best}}$.
  \end{algorithmic}
\end{algorithm}

\section{NUMERICAL RESULTS}
\label{sec 4}

\subsection{Simulation Test}

In simulations, the target BS is randomly placed ranging from $100$ to $300$ meters from the DMA of the relay, while $4$ interference BSs are placed $300$ to $900$ meters away.
All results are averaged over $500$ independent simulations of randomly generated BS positions. 
For simplicity, all BSs have the same transmission power, which is set to $1$.

The DMA is 2-bit quantized and contains a $4\times4$ antenna array, with an element spacing of $\frac{\lambda}{2}$. The initial population size is $100$, and the maximum number of iterations $T_{max}$ is $50$. 
We set the rotation angle step size $\triangle\omega_{0}$ to $0.01\pi$, the block size parameter $b$ to $2$, the rotation angle weighting factor $a$ to $0.6$, and a basic mutation probability $P_{0}$ to $0.1$.
Apart from DBQG, several baseline schemes are also evaluated:
\begin{itemize}
    \item Random-Max Algorithm (RMA): Randomly generate phase coefficients and select the coefficients that yield the optimal gain in performance metrics as the final result.
    \item MMSE: Utilize the minimum mean square error criterion for receiver-side filtering. 
    However, the weighted matrix needs to be quantized on discrete DMA, which results in performance degradation.
    \item Greedy fast beamforming algorithm (GFBA) \cite{RIS_pei}: Gradually changes the phase shifts column-by-column or row-by-row while maintaining the optimal phase shift coefficients, thereby gradually improving performance metrics.
\end{itemize}

\begin{figure}[htbp]
\centerline{\includegraphics[width = .62\linewidth]{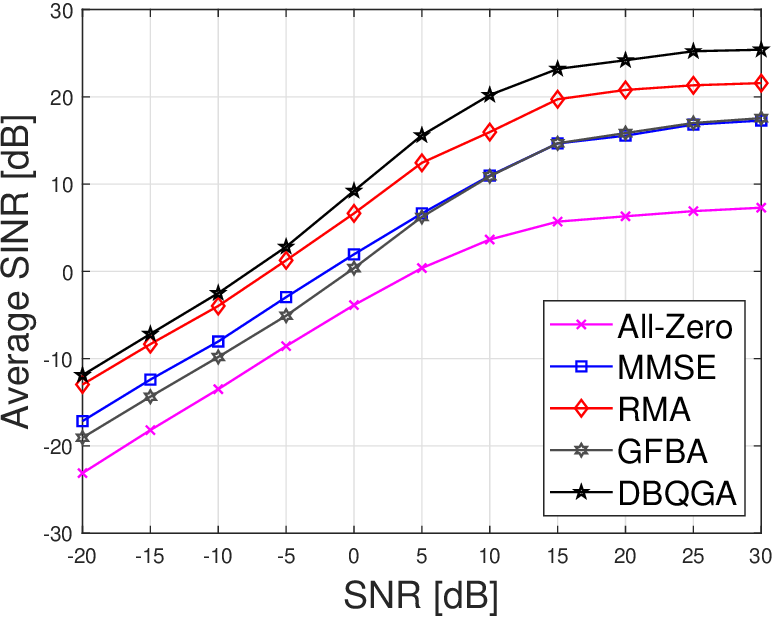}}
\caption{SINR of tested algorithms vs. the SNR of All-Zero DMA.}
\label{fig1}
\end{figure}

In the simulations, the number of random states for RMA was set to $5000$, which is equal to the amount of feedback when simulating the DBQG method. Fig.\ref{fig1} illustrates the average SINR of each algorithm under different noise conditions. 
The horizontal axis of the figure represents the SNR when the DMA phase shift coefficients are all set to $0$ under various noise conditions.
It can be observed that DBQG demonstrates more significant SINR improvement than other algorithms. 
Specifically, under lower noise conditions, DBQG achieves a performance improvement of $4$ dB compared with RMA at the same complexity, while the performance improvement over GFBA is even more remarkable.

\begin{figure}[htbp]
\centerline{\includegraphics[width = .62\linewidth]{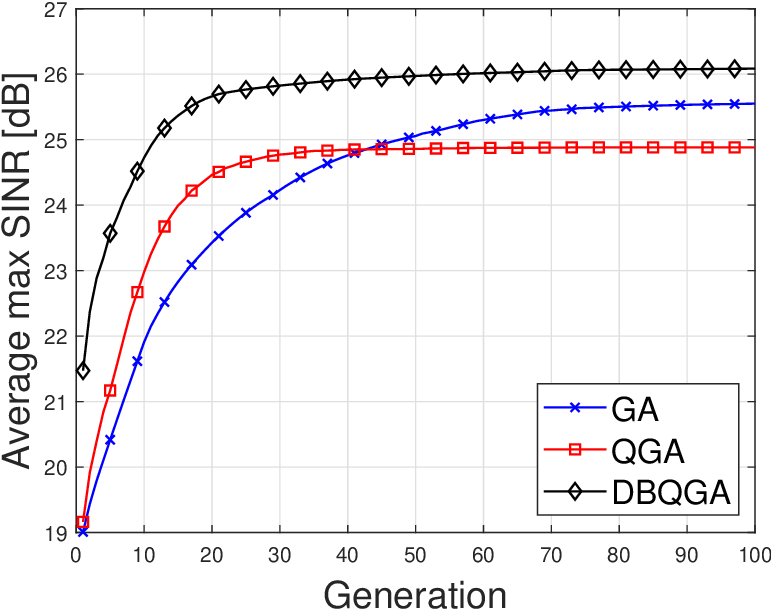}}
\caption{Average maximum SINR per generation of the algorithms.}
\label{fig2}
\end{figure}

In addition, comparisons were made between DBQG and the classical GA
as well as the QGA. 
For better comparison, every algorithm was evaluated for $500$ trials under the same conditions, and the average values were plotted in Fig.\ref{fig2}.
Due to the implementation of the dynamic block initialization strategy, DBQG possesses a higher initial value.
Additionally, the convergence speed of DBQGA is about $50\%$ faster than classic GA. Since the complexity of each generation is $\mathcal{O}(G)$ in both GA and DBQGA, the overall computational complexity of the proposed DBQGA is reduced by $50\%$.
Notably, when addressing this problem, QGA exhibits a significant local convergence phenomenon, whereas DBQG successfully resolves this issue and improves the final optimization result.

\subsection{Field Test}

In this section, we present the outdoor field trials in the commercial 5G network of China Mobile when deploying DBQG on an AF relay system equipped with DMA. 
The DMA  is a $4\times4$ uniform planar array with $2$-bit quantization.
The distance between the target BS and the DMA is approximately 200 meters.
During the test, severe inter-cell interference exists from more than 5 BSs in the 2.6 GHz N41 frequency band.

\begin{figure*}[htbp]
\begin{minipage}[b]{0.31\textwidth}
    \centering
    \centerline{\includegraphics[width = \linewidth]{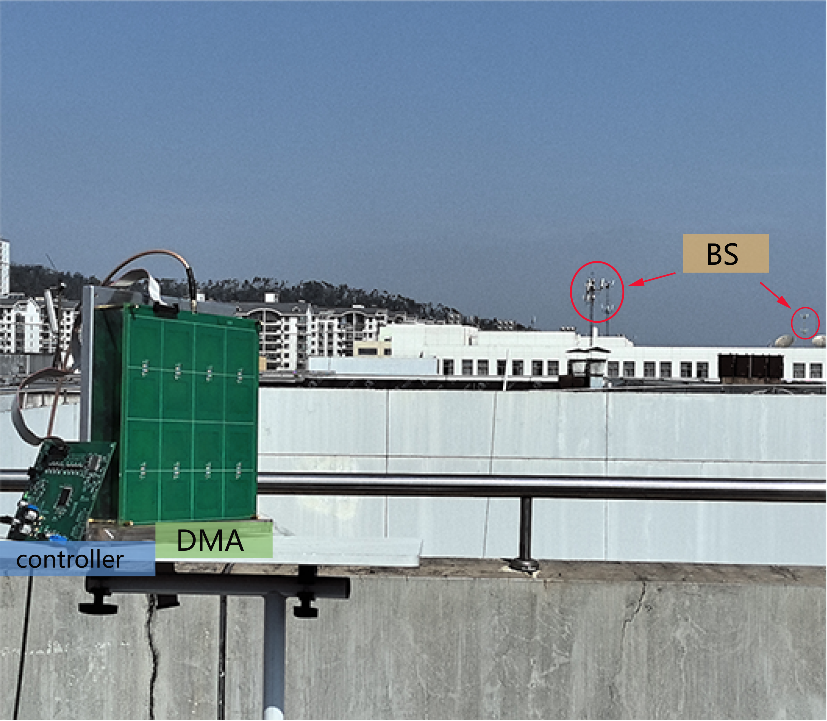}}
    \caption{Field test scenario.}
    \label{fig:Field_test_scenario}
\end{minipage}
\hfill
\centering
\begin{minipage}[b]{0.31\textwidth}
    \centering
    \includegraphics[width=\linewidth]{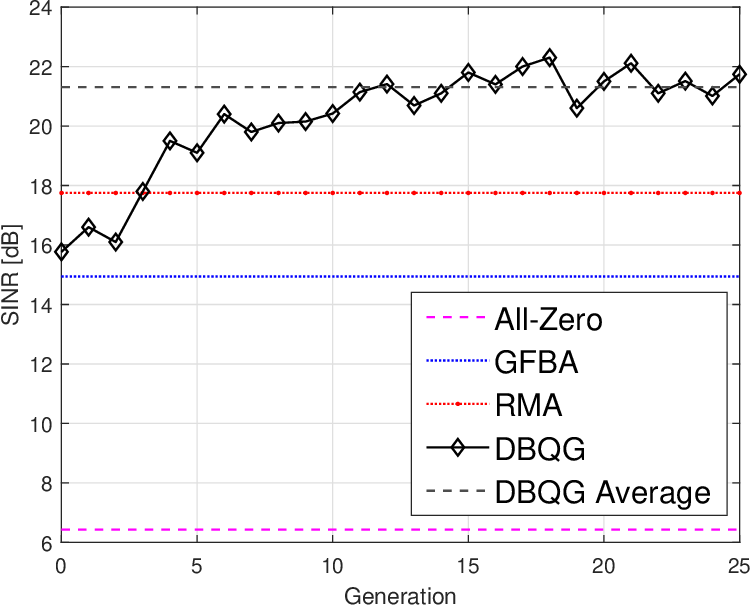}
    \caption{Field test maximum SINR per generation in setup 1.}
    \label{field_test_result2}
\end{minipage}
\hfill
\begin{minipage}[b]{0.31\textwidth}
    \centering
    \includegraphics[width=\linewidth]{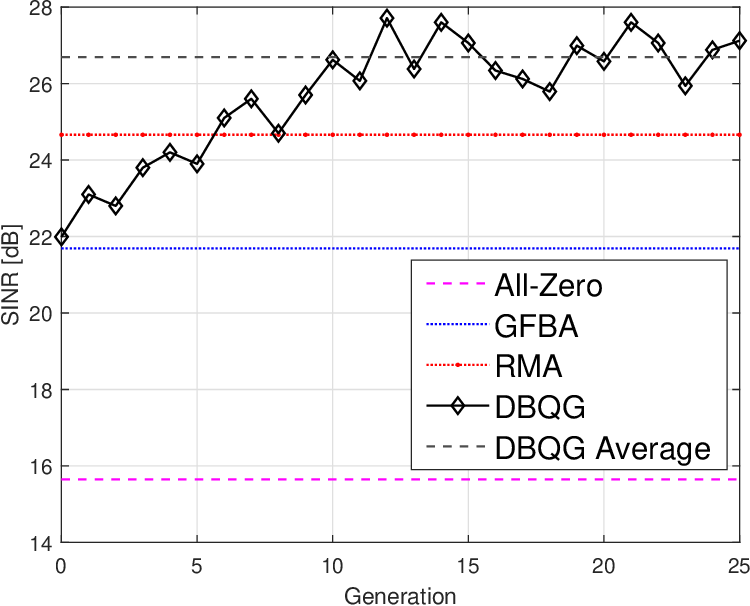}
    \caption{Field test maximum SINR per generation in setup 2.}
    \label{field_test_result3}
\end{minipage}
\end{figure*}

The photo of the experimental scenario is shown in Fig.\ref{fig:Field_test_scenario}. 
To bring our prototype system closer to real-world scenarios, we obtain signal metrics from a 5G test phone and feed them back to the DMA control board via Bluetooth.
Throughout the testing process, no knowledge about the BS side was assumed, and no coordination from the BS side was required. 

In the experiment, the parameters of DBQG were consistent with those used in the simulation mentioned above. Alongside DBQG, the experiments are also made with the GFBA, as well as the RMA with 2000 random states for fair comparison. 
The tests were conducted under three different setups (varying DMA angles and locations). 
During the field test, we found that the actual wireless environment is highly volatile, necessitating the averaging of the obtained signal metrics.

The final average SINR results are shown in Table \ref{tab:field_test_sinr}. Fig.\ref{field_test_result2} and Fig.\ref{field_test_result3} illustrate the variation trend of the maximum SINR over the iterations for the DMA with DBQG deployed in two test cases. 
The final average SINR values of GFBA and RMA under the same conditions are added for comparison. 
It is noted that the maximum SINR of DBQG in the field test is not monotonically increasing due to the severe fluctuation of 
wireless signals in the real communication environment.

\begin{table}[htbp]
\caption{Field Test Results of Final Average SINR}
\begin{center}
\resizebox{.6\linewidth}{!}{ 
\begin{tabular}{|c|c|c|c|c|}
\hline
\textbf{Test}&\multicolumn{4}{c|}{\textbf{Tested Algorithms (unit: dB)}} \\
\cline{2-5} 
\textbf{setups} & All-Zero & GFBA & RMA & DBQG \\ 
\hline
1    & 6.43     & 14.94  & 17.75  & 21.31  \\ \hline
2    & 15.65    & 21.69  & 24.66  & 26.69  \\ \hline
3    & 11.23    & 16.81   & 18.21  & 20.79  \\ \hline
\end{tabular}
}
\label{tab:field_test_sinr}
\end{center}
\end{table}

During the test, it was found that after around the $15$-th generation, the average SINR no longer increases, leading to less SINR improvement than observed in simulations. 
It is currently believed that the significant fluctuations in the wireless environment, coupled with the limited accuracy of the equipment, result in errors in the signal indicators obtained by the controller, thereby affecting the optimization performance of the algorithm. 
Additionally, the phase adjustment of the DMA is not as precise as in the simulation, which is also one of the contributing factors.

\section{CONCLUSION}

In this paper, we designed a quantum genetic algorithm that introduces an efficient dynamic block initialization and overarching block update strategy that does not require any channel information to suppress interference in a DMA relay system.
Simulations demonstrate significant performance gains and high convergence speed of the algorithm. 
Outdoor field tests with our self-built DMA relay system in commercial 5G networks validate the practicality of the algorithm and also demonstrate the capability of DMA to address interference suppression problems in real-world communication networks.

\bibliographystyle{IEEEtran}
\bibliography{IEEEabrv, references.bib}

\end{document}